\documentclass[iop]{emulateapj}

\usepackage{amssymb, amsmath, natbib, graphicx, multirow}
\usepackage{xspace, ulem}
\usepackage{relsize, isotope}

\newcommand{\coo}{CO$_2$\xspace}
\newcommand{\hoh}{H$_2$O\xspace}
\renewcommand{\micron}{$\mu$m\xspace}

\shorttitle{HR 4049: The Mid-IR Spectrum}
\shortauthors{Malek \& Cami}
\slugcomment{Received May 21 2013; accepted October 19 2013}

\begin{document}
\title{The Gas-Rich Circumbinary Disk of HR~4049 I: A Detailed Study
of the Mid-Infrared Spectrum.}

\author{S.E.~Malek\altaffilmark{1} and J.~Cami\altaffilmark{1,2}}
\altaffiltext{1}{Department of Physics \& Astronomy, University of Western
  Ontario, London, ON N6A 3K7, Canada}
\altaffiltext{2}{SETI Institute, 189 Bernardo Avenue, Suite 100,
  Mountain View, CA 94034, USA}
\email{sarahemalek@gmail.com}
\email{jcami@uwo.ca}

\normalem

\begin{abstract}
We present a detailed analysis of the mid-infrared spectrum of the
peculiar evolved object HR~4049. The full Spitzer-IRS high-resolution
spectrum shows a wealth of emission with prominent features from \coo
and \hoh and possible contributions from HCN and OH. We model
the molecular emission and find that it originates from a massive ($M
\gtrsim 8 \times 10^{-3}$ M$_{\sun}$), warm ($T_{\rm ex} \approx 500$~K) and
radially extended gas disk that is optically thick at infrared
wavelengths. We also report less enrichment in $^{17}$O and $^{18}$O
than previously found and a comparison of the Spitzer observations to
earlier data obtained by ISO-SWS reveals that the \coo flux has more
than doubled in 10 years time, indicating
active and ongoing chemical evolution in the circumbinary disk. If the
gas originates from interaction between the stellar wind and the dust,
this suggests that the dust could be oxygen-rich
in nature. The molecular gas plays a
crucial role in the thermal properties of the circumbinary disk by
allowing visible light to heat the dust and then trapping the infrared
photons emitted by the dust. This 
results in higher temperatures and a more homogeneous temperature
structure in the disk.
\end{abstract}


\section{Introduction}

\object{HR 4049} is considered the prototype for a class of evolved objects
with peculiar properties. Their effective temperatures and
luminosities suggest that they are in the post-asymptotic giant branch
(post-AGB) phase of their evolution, but their evolutionary path may be
severely affected by the presence of a companion \citep[see][for a
review]{VanWinckel1995}. Indeed, their unusual properties all seem to
result from stellar evolution in a binary system.

Like many of the members of this class, HR 4049 shows a significant
infrared (IR) excess and ultraviolet (UV) deficit \citep{Lamers1986},
suggesting the presence of a massive circumbinary disk. This disk is
the result of mass loss in the binary system and plays a significant
role in its unusual properties. For instance, the photospheric
abundances of HR~4049 show an extreme depletion in refractory elements
\citep[e.g.~{[Fe/H]}~=~-4.8,][]{Waelkens1991} while showing nearly
solar abundances for volatiles \citep[e.g.~{[S/H]}~=~-0.2,
{[C/H]}~=~-0.2, {[N/H]}~=0.0, {[O/H]}~=~-0.3 ][]{TakadaHidai1990,
Waelkens1991}. This peculiar depletion pattern is the result of dust
formation in the disk followed by accretion of the gas which is now
devoid of refractory elements \citep{Mathis1992, Waters1992}. The
circumbinary disk also causes HR 4049 to display photometric
variability which is tied to its orbital period \citep{Waelkens1991}.

Due to its importance in defining the characteristics of HR 4049, the
dust component of the circumbinary disk has been the subject of a
number of studies \citep[e.g.][]{Waelkens1991a,Dominik2003,Acke2013},
revealing unusual dust properties. The infrared spectrum of HR~4049
does not show the dust features typically associated with
circumstellar material around evolved stars (e.g.~silicates or SiC)
and thus the nature of the dust remains a mystery. Instead, the
infrared excess is well-represented by a single temperature ($T\approx
1150$~K) black body from the onset of dust emission in the near-IR to
submillimetre (submm) wavelengths. At the same time, the dust re-emits
a significant fraction of the total stellar luminosity ($L_{\rm IR}
\approx L_* / 3$), from which \citet{Dominik2003} deduced that 
HR 4049 is surrounded by an extremely optically thick
and vertically extended disk with a hot inner wall. In this so-called
``wall model", the inner rim of the disk is 10 AU from the center of
the binary system with a scale height of 3~AU and a temperature of
1150 $\pm$ 150 K. 

However, \citet{Acke2013} found that the wall model was inconsistent
with interferometric observations of HR~4049 which showed a radially
extended disk. Instead, they proposed an optically thin dust disk
composed of minerals without strong dust features---amorphous carbon
being the most probable. In their model, the dust is slightly further
from the center of the system and extends to much larger scale
heights. Their model does not fit the spectral energy distribution
(SED) beyond 20~\micron, so they included a dust component composed of
large grains at 200~K. 

One potential clue to the nature of the dust (and the evolutionary status of the
system) is the presence of strong emission features due to
polycyclic aromatic hydrocarbons \citep[PAHs, see][]{Waters1989,
Tielens2008} as well as from nano-diamonds at 3.43 and 3.53~\micron
\citep{Geballe1989, Guillois1999} and C$_{60}$ \citep{Roberts2012}.
Such species are often found in the environments surrounding evolved,
carbon-rich objects, which suggests that HR~4049 may have once been a 
carbon star. 

However, the IR spectrum also reveals a plethora of molecular bands from
species that are more characteristic of oxygen-rich environments, but
are very unusual nonetheless: \citet{Cami2001} detected and identified
the emission features due to all possible isotopologues of \coo
containing \isotope[13]{C}, \isotope[17]{O} and \isotope[18]{O}. Using
optically thin models, they found that the gas was extremely enriched
in \isotope[17]{O} and \isotope[18]{O}
(with\isotope[16]{O}/\isotope[17]{O} = 8.3 $\pm$ 2.3 and
\isotope[16]{O}/\isotope[18]{O} = 6.9 $\pm$ 0.9). 
Subsequently, \citet{Hinkle2007} analyzed the first overtone and
fundamental bands of CO in the near-IR spectrum and found no
enrichment in \isotope[17]{O} and \isotope[18]{O}. Although they
detected CO isotopologues containing \isotope[17]{O} and
\isotope[18]{O} in the fundamental, they did not observe these
isotopologues in the overtone. They attributed the discrepancy between
their oxygen abundances and those detected by \citet{Cami2001} to the
optically thick nature of the \coo emission in the mid-IR
observations. 

\medskip

A detailed study of the gas has been difficult thus far due to the low
sensitivity of the observations taken with the Short Wavelength
Spectrometer on board the Infrared Space Observatory
\citep[ISO-SWS,][]{deGraauw1996} as well as limitations in earlier
line lists describing the transition energies and probabilities of
the molecular species present in this spectrum. However, a good
understanding of the gas composition should yield clues to the
evolutionary history of this system as well as to the processing that
occurs in these environments.

Here, we present observations from the Infrared Spectrograph
\citep[IRS,][]{Houck2004} on board the {\em
Spitzer} Space Telescope
\citep{Werner2004} which have a higher signal-to-noise (S/N) ratio as
well as better spectral resolution. In addition,
due to improvements in available line lists, we will be able to
include the effects of optical depth in our spectral models.

In addition to the \coo bands, we analyze the other spectral
features in the Spitzer-IRS and ISO-SWS spectra. We provide an
inventory of the gas components and compare the observations to
molecular model spectra to obtain excitation temperatures, column
densities and optical depths. In turn, these numbers provide
quantitative information on the gas disk which we compare to earlier
studies of the dust in this object. We then compare our results from
these analyses to the CO observations in the near-IR spectrum from
Phoenix on Gemini in Malek \& Cami (submitted, hereafter Paper
II). 
%\citet[hereafter Paper II]{Malek2013a}.

This paper is organized as follows. In Section~\ref{sec:datareduction},
we describe the observational data and reduction steps. We describe
our modeling technique in Section \ref{sec:analysis} and present our
results of this analysis in Section \ref{sec:results}. Then we
describe how these results fit with the other observations of
the system in Section \ref{sec:discussion} and finally, we present our
conclusions in Section \ref{sec:conclusion}.

\begin{figure}[t!]
\includegraphics[width=0.48\textwidth]{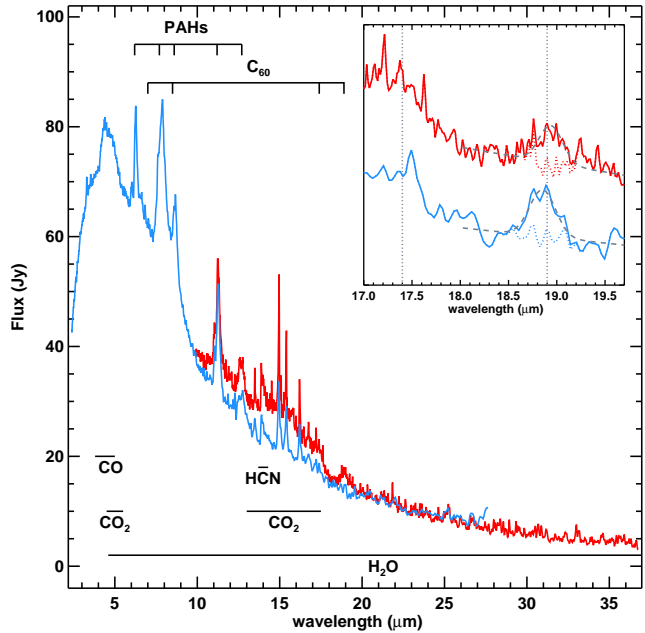}
\caption{The Spitzer-IRS (black, red in the online version) and
speed 2 ISO-SWS (light gray, blue in the online version) observations
of HR 4049 are shown.  The polycyclic aromatic hydrocarbon (PAH)
  bands at 6.2, 7.7, 8.6, 11.2 and 12.7 $\mu$m are highlighted as well
  as the positions of the C$_{60}$ bands at 7.0, 8.5, 17.4 and 18.9
  $\mu$m.  The regions of the spectrum showing emission from CO, \coo,
  \hoh, and HCN are also highlighted.  The inset shows the 17 to 19.5
  $\mu$m region of the Spitzer-IRS (black, red in the online version)
and speed 1 ISO-SWS spectra (light grey, blue in the online version)
highlighting the locations of the 17.4 and 18.9 $\mu$m
  C$_{60}$ bands and showing the best-fit Gaussians to the 18.9 $\mu$m
  bands (dashed gray lines) with the corresponding residuals for the
  fitting region (dotted lines).} 
\label{fig:midobservations}
\end{figure}

\section{Observations and Data  Reduction}
\label{sec:datareduction}

\subsection{Spitzer-IRS}

HR 4049 was observed at high spectral resolution ($R = 600$)
using the Infrared Spectrograph \citep[IRS,][]{Houck2004} aboard the
Spitzer Space Telescope \citep{Werner2004} in the short high (SH,
$\lambda$ = 9.9 to 19.6 $\mu$m) mode (AOR key 4900608; program ID 93; PI
D. Cruikshank) and the long high (LH, $\lambda$ = 18.7 to 37.2 $\mu$m)
mode (AOR key 23711232; program ID 40896; PI J. Cami). For the LH
data, we also obtained background observations (AOR key 23711488).

We used the {\sc smart} data reduction package
\citep[v8.2.2,][]{Hidgon2004} as well as custom IDL routines to carry
out the data reduction.  Starting from the basic calibrated data
(bcd), we first cleaned all data with {\sc irsclean} using the
campaign rogue pixel mask. When we examined the LH background observations,
we found the background flux levels were fairly low (typically less than
1\% of the target flux with some small unresolved spikes up to 5\% of
the target flux) and featureless. We thus subtracted this background
from the target observations. We then combined the cleaned data
collection events for each observation using a weighted average. Next,
we extracted the spectra from the SH data and the background
subtracted LH data using full aperture extraction in {\sc smart}
v8.2.2 \citep{Hidgon2004}. 

We then defringed the extracted spectra and trimmed the edges of the
orders. Flux differences between adjacent orders were typically on the
order of 5\% among the SH orders and only one LH order showed a
notable flux difference.  We scaled adjacent orders to the median flux
in the overlap region, using order 11 ($\lambda_c$ = 18.7 \micron for
SH and $\lambda_c$ = 35.4 \micron for LH) as the reference. We then
compared the spectra in overlap regions as well as the two nod
positions while checking for consistency, averaged them using a
weighted mean, and rebinned the resulting final spectrum. 

We compared the long wavelength end of the SH spectrum and the short
wavelength end of the LH spectrum and calculated a weighted mean for
the overlap region.  Finally, we scaled the Spitzer-IRS spectrum to
the Infrared Astronomical Satellite (IRAS) flux measurement at 25
\micron (scaling up by 50\%). %\footnote{We also looked at the
%WISE data, but discovered that HR 4049 was saturated in all but the W4
%bandpass, which was consistent with the IRAS 25 \micron flux.}  
The final spectrum is presented in Fig.~\ref{fig:midobservations}. 

\subsection{ISO-SWS}

HR 4049 was observed twice with the Short Wavelength Spectrometer
\citep[SWS,][]{deGraauw1996} aboard the Infrared Space Observatory
\citep[ISO,][]{Kessler1996}: on December 27th, 1995 (AOT 1, speed 1)
and again on May 6th, 1996 (AOT 1, speed 2); both observations
correspond to a resolving power of $\sim$300. Here, we use the speed
2 ISO-SWS data that was also presented by \citet{Cami2001} and
\citet{Dominik2003}. 

We compare the speed 2 ISO-SWS spectrum to the Spitzer-IRS
observations in Fig.~\ref{fig:midobservations}. 
At the longer wavelengths, the spectra agree well
with one another; however, the Spitzer-IRS data exhibit a significant
increase in the flux levels between 10 and 18 $\mu$m. 

\begin{figure*}[t!]
\centering
\includegraphics[width=1\textwidth]{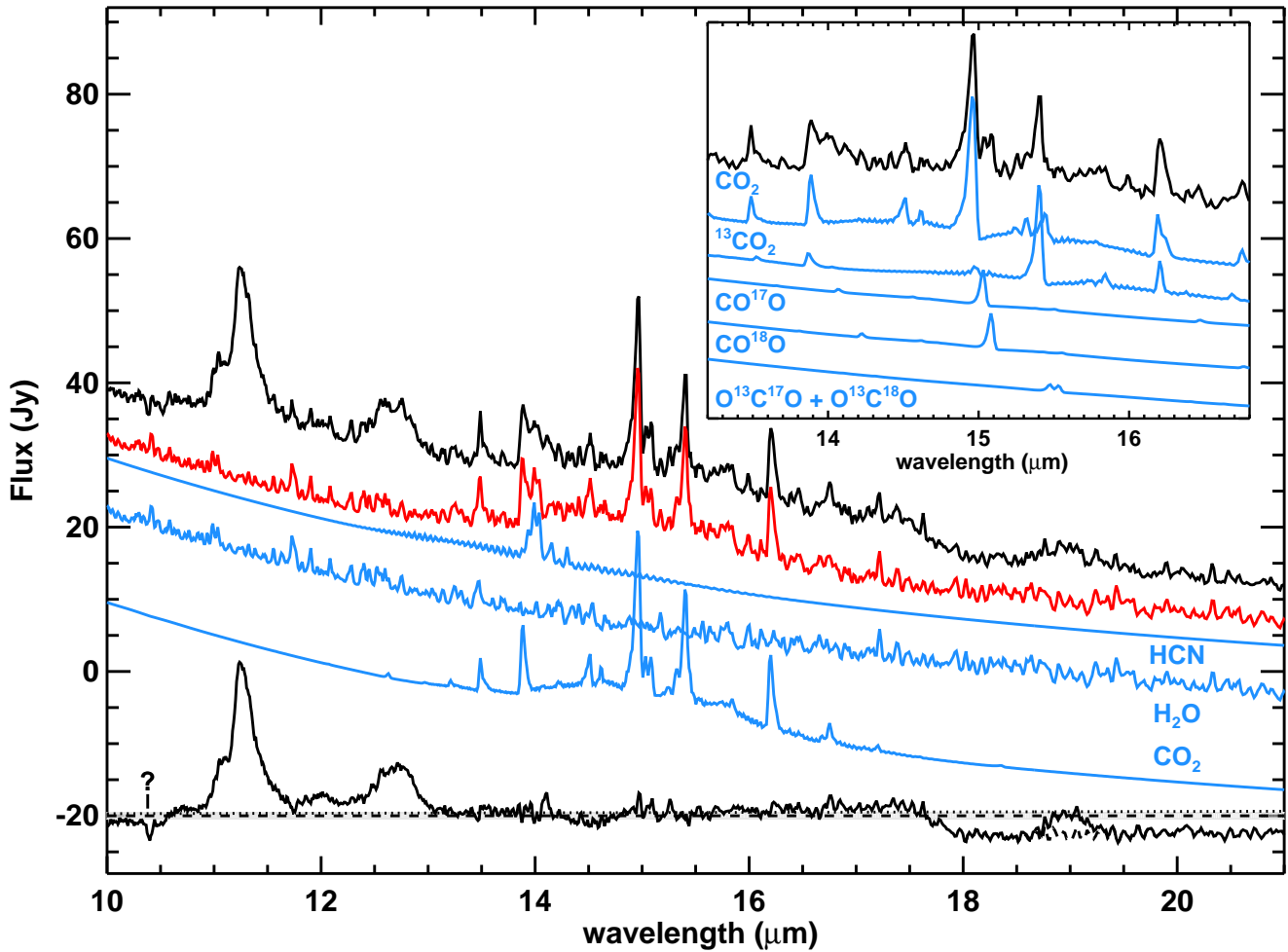}
\caption{The Spitzer-IRS spectrum between 10 and 20 \micron (black)
  with the best fit model for all the molecules (dark gray or red in
  the online version) offset below the spectrum and models of
  the individual molecules below that (gray or blue in the online
  version) in descending order: HCN, \hoh, and \coo.  The residual
  without the linear component is shown at the bottom of the figure in
  black with the error values on the data indicated in light gray. We
  also fit a Gaussian to the C$_{60}$ feature at 19 \micron in the
  residual and show the effect of removing this Gaussian in dashed
  lines. The inset shows a breakdown of the \coo isotopologues between
  13 and 17 \micron.}
\label{fig:sh_zoom}
\end{figure*}

\begin{deluxetable}{l c c}
\tablecolumns{3}
\tablewidth{0pt}
  \tablecaption{Temperatures, column densities, and isotope ratios
   for the best model fits to the data as well as the nominal
3$\sigma$ uncertainties. }
\startdata
\hline \hline \\
 & Spitzer-IRS & ISO-SWS \\ [1.1ex] 
$\lambda_{\rm fit}$ (\micron) & 13.7--18 & 2.4--5.8; 9.5--10.5; \\
                              &          &  14--22 \\ \hline
\\ [1.1ex] 
Temperature (K) & 500 $\pm$ 50 &   600 $\pm$ 50 \\ [1.1ex]
log $N$ (\coo)  & 19.0 $\pm$ 0.1  &   17.8 $\pm$ 0.1 \\ [1.1ex]
log $N$ (\hoh)  & 21.6$^{+0.1}_{-0.2}$  &   19.4 $\pm$ 0.1 \\ [1.1ex]
log $N$ (HCN)   & 17.8 $\pm$ 0.1 & ... \\ [1.2ex]
log $N$ (CO)    & ... & 22.0 $\pm$ 0.1 \\ [1.2ex]
\isotope[12]{C}/\isotope[13]{C} & 6$^{+2}_{-1}$ &  1.6$^{+2}_{-0.4}$ \\ [1.2ex]
\isotope[16]{O}/\isotope[17]{O} & 160$^{+90}_{-35}$ & 40$^{+23}_{-9}$ \\ [1.2ex]
\isotope[16]{O}/\isotope[18]{O} & 160$^{+40}_{-60}$ & 16$^{+4}_{-3}$ \\ [1.2ex]
\isotope[14]{N}/\isotope[15]{N} & 13$^{+3}_{-8}$ & ... \\ [1.2ex]
\enddata
\label{table:mirfits}
\end{deluxetable}

\section{Analysis}
\label{sec:analysis}

Fig.~\ref{fig:midobservations} shows the full Spitzer-IRS spectrum
with emission from a variety of molecular species indicated. There are
prominent features from large molecules such as polycyclic
aromatic hydrocarbons \citep[PAHs,][]{Waters1989} and C$_{60}$
\citep{Roberts2012}.  We have also indicated the broad
emission features from \coo at 15~\micron, the HCN feature at
14~\micron and the region of the spectrum where we observe \hoh
emission. There is also emission from CO and \coo at 4.6 and
4.2~\micron respectively in the ISO-SWS spectrum.

\subsection{Modeling the Spitzer Spectrum}

To determine the properties of the gas in the mid-IR spectrum
of HR~4049 and fully characterize the molecular emission, we created
model spectra and compared them to the observational data. We used the
same methods employed to build the SpectraFactory database
\citep{Cami2010a}. For each model, we began with line lists detailing
the frequencies and intensities of individual molecular transitions.
We calculated optical depth profiles from the line lists assuming a
population in local thermodynamic equilibrium (LTE) and a Gaussian
intrinsic line profile with a width of 3 km s$^{-1}$.  We summed the
optical depth profiles for the different molecular species (including
isotopologues) and then performed the proper radiative transfer
calculations through an isothermal slab and smoothed the resulting
model spectrum to match the SH resolution ($R = 600$).

We used a 1150 K black body for the continuum and applied a non-negative
least-squares (NNLS) algorithm \citep{Lawson1974} with the continuum
and the molecular emission models as parameters. Then we compared our
model to the observational data between 13.27 and 18 \micron (to cover
the \coo emission) and calculated $\chi^2_{\nu}$, the reduced $\chi^2$
statistic for each model to determine the quality of the fit. 

We experimented with several different models to determine the best
range for each of our parameters then selected the parameters for each model
using an adaptive mesh algorithm. In our final fits, we varied the
temperature of the molecular layer between 200 and 1000 K in
increments of 100 K and column densities between 10$^{16}$ and
10$^{22}$ cm$^{-2}$ in increments of log$N$ = 0.2 for all molecules in
this region. Additionally, we varied log($^{12}$C/$^{13}$C) from 0 to
2; log($^{16}$O/$^{18}$O), log($^{16}$O/$^{17}$O), and
log($^{14}$N/$^{15}$N) from 0 to 3 in increments of 0.2. 

While we began fitting only \coo and its isotopologues, we later
included \hoh and HCN in our model spectra. In addition, we noted a
small linear trend in our residuals from 13 to 17.5~\micron. We do not
know the origin of this trend, but we incorporated a linear component
in our NNLS routine to compensate for this residual.

\section{Results}
\label{sec:results}

\begin{figure}[t!]
\centering
\includegraphics[width=0.5\textwidth]{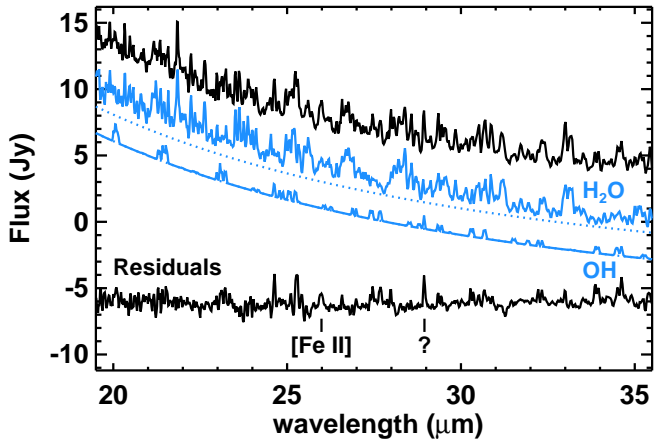}
\caption{The Spitzer-IRS spectrum of HR 4049 (black) with the best fit
model for \hoh and a model for OH at the same temperature and column
density offset below the spectrum (gray or blue in the online
version). The dotted gray (blue in the online version) lines indicate
the black body continuum under the \hoh and OH models.}
\label{fig:spitzeroh}
\end{figure}

We present the parameters for our best fit model in Table
\ref{table:mirfits} and the 10 to 20 \micron region of the Spitzer-IRS
spectrum with our best fit model in Fig.~\ref{fig:sh_zoom}. We focus
on the \hoh emission at LH wavelengths in Fig.~\ref{fig:spitzeroh} and
we compare the models to the full spectrum Spitzer-IRS and our
predictions at ISO-SWS wavelengths in Fig.~\ref{fig:midir_fits}. We
find a $\chi^2_\nu$ of 3.5 for the fit to the Spitzer-IRS
observation and a good representation of the molecular
emission features within our fitting region.

When we compare the predictions from our model at longer and shorter
wavelengths to the spectrum, we find that the spectral features are
also fit remarkably well. The majority of the molecular features in
the LH spectrum, for example, appear to be from \hoh emission. In
addition, the spikes on the PAH features at 11.2 and 12.7~\micron
disappear and some of the smaller PAH features become evident. 

\subsection{\coo}

The Spitzer-IRS spectrum shows prominent emission from the \coo
isotopologues observed by \citet{Cami2001} in addition to others
(e.g.~the O\isotope[13]{C}\isotope[17]{O} and
O\isotope[13]{C}\isotope[18]{O} peaks which \citet{Cami2001} were
unable to separate due to the lower spectral resolution of ISO-SWS).
To model the emission from \coo, we used line lists from the 1000~K
Carbon Dioxide Spectroscopic Database \citep[CDSD,][]{Tashkun2003}
including the $^{13}$CO$_2$, CO$^{17}$O, CO$^{18}$O,
O$^{13}$C$^{17}$O, and O$^{13}$C$^{18}$O isotopologues. 

Though our single layer LTE model reproduces the majority of the \coo
features very well, (see Fig.~\ref{fig:sh_zoom}), we find that the
best fit model tends to slightly overestimate the flux relative to the
local continuum between 14.2 and 14.7~$\mu$m while underestimating the
flux between 16.7 and 17.6~$\mu$m (which may be due to some
contribution from C$_{60}$ at the long wavelength end). We also find
high optical depths across most of our fit region with this model. 

There are also small spikes which may be due to a
slight temperature stratification in the \coo layer. For example, we
note a small feature in the residual 
spectrum at 16.64~\micron which could be due to the transition
between the $4\nu_2^4$ and $3\nu_2^1$ levels, which suggests the
presence of additional hotter gas. We also observe a small residual
peak at 14.98~\micron near the main \coo band, which is a combination
of the $\nu_2$ bending mode at 14.98~\micron ($1\nu_2^1$ to the ground
state) and subsequent hot bands which are each shifted slightly to the
blue; the presence of residual emission at 14.98~\micron thus suggests
the presence of some colder \coo. The high optical depths of the gas
could make the appearance of the bands more sensitive to these types
of temperature variations; however, these residuals are relatively
small, suggesting that most of the emission originates from a
relatively thermally homogeneous layer.

We also note some small residual emission at 15.04 and 15.10~\micron,
which could be due to additional emission from the main $\nu_2$ bands of
the OC\isotope[17]{O} and OC\isotope[18]O isotopologues respectively.

\subsection{\hoh and OH}

After including all \coo isotopologues in our models, many weaker
emission features remained in the residuals of this region. We
noticed that several features are consistent with emission from water
vapor. Thus, we recalculated our models including optical depth 
profiles for \hoh using line lists from \citet[including the 
H$_2^{17}$O and H$_2^{18}$O isotopologues]{Partridge1997}.

We find an extremely high column density for \hoh and a much better
correspondence to the results than for \coo, not only in the region we
chose to fit, but also at longer and shorter wavelengths in the
Spitzer-IRS spectrum. Upon careful examination of the residuals in
Fig.~\ref{fig:sh_zoom}, one may note that the spikes atop the 11.2 and
12.7 \micron PAH features disappear almost completely and remarkably,
the 12 \micron PAH feature becomes apparent though it was previously
hidden by \hoh emission. Furthermore, the 18.9~\micron C$_{60}$
feature becomes much more prominent and clear when \hoh emission is
removed. In Figs.~\ref{fig:spitzeroh} and \ref{fig:midir_fits}, it is
also clear that \hoh accounts for the bulk of the features in the LH
spectrum since the residuals contain relatively few remaining spectral
features. This observation also confirms the earlier detection of
water in HR~4049 in the near-IR by \citet{Hinkle2007}.  

\medskip

Since OH was also detected by \citet{Hinkle2007} at $\sim$3~\micron,
and since it has many features in this wavelength range, we
included it in our models between 13.24 and 18 \micron. However, we
were unable to detect OH in this region of the spectrum. We were also
unable to fit any OH using only \hoh and OH between 20 and 35 \micron.
In Fig.~\ref{fig:spitzeroh}, we show a comparison between model
spectra of \hoh and OH using the same temperatures and column
densities at LH wavelengths. We note that while there appear to be
many features in our residuals which are consistent with emission from
OH, it is not possible to reliably fit the OH features due to
extensive contamination from \hoh.

\subsection{HCN}

When all the \coo and \hoh isotopologues are included in our models,
there is still significant residual emission at 14.04~\micron, where the
CH bending mode of HCN is often seen in evolved stars. Indeed, the
SpectraFactory catalogue \citep{Cami2010a} reveals a clear HCN
molecular band at this wavelength, thus we include it in our model
calculations using line lists from the HITRAN 2008 database
\citep[including the H$^{13}$CN and HC$^{15}$N
isotopologues]{Rothman2009}.

We determine a column density for HCN which is much lower than that of
\coo and \hoh (log$N = 17.8$), which suggests that it is less
abundant. We included isotopologues of HCN containing \isotope[13]{C}
and \isotope[15]{N} which appear slightly to the red of the main
isotopologue and are thus able to find \isotope[14]{N}/\isotope[15]{N}
of 13$^{+3}_{-5}$. With only one band containing this isotope though,
we do not consider our ratio here to be especially reliable.

\subsection{Fullerenes}

Since its detection in the young planetary nebula Tc~1 by
\citet{Cami2010}, C$_{60}$ has been reported in other evolved binary
systems \citep{Gielen2011} and it was recently reported in HR 4049 by
\citet{Roberts2012}, who detected the 17.4 and 18.9~\micron features. 

Indeed, the 18.9 \micron feature is clear (and also present in the
ISO-SWS speed 1 observations, see the inset of
Fig.~\ref{fig:midobservations}) and when we subtract our best fit \hoh
model, the feature becomes much more prominent (see
Fig.~\ref{fig:sh_zoom}). Fitting the 18.9 \micron C$_{60}$ feature in
the residual spectrum with a Gaussian, we determine a
FWHM of 0.40~\micron and a central wavelength of 18.98~\micron.  This
is a much narrower band than found by \citet{Roberts2012}, who
reported a FWHM of 0.64~\micron. This difference may be due to the
presence of the water features, which makes the band appear
broader.

The 17.4 \micron feature is buried in the optically thick \coo
emission, so we cannot measure it accurately.  Additionally, since no
observations were taken by the IRS short low (SL; $R=90$ to $127$,
$\lambda = 5.2$ to $14.5$ \micron) mode, the 7 and 8.5~\micron
features for C$_{60}$ would only appear in the ISO-SWS spectrum.
However, we do not see them, which could be due to a combination of
the sensitivity of ISO-SWS and the often weak nature of these
features. 

\subsection{PAHs}
\label{sec:results_pahs}

The spectral region covered by the Spitzer-IRS observations covers
only the 11.2 and 12.7~\micron PAH features. These bands also appear
in the ISO-SWS spectrum alongside features at 6.2, 7.7 and
8.6~\micron. The PAH features in the ISO-SWS spectrum of HR 4049 were
described in detail by \citet{Beintema1996} and \citet{Molster1996}
and are class B \citep{Peeters2002}.

Examining the residual spectra (see Fig.~\ref{fig:sh_zoom}
especially), we note that some of the less prominent PAH features
become clear when the \hoh emission is removed from the spectrum. 
For instance, the 12 \micron feature from CH out-of-plane duo-modes
and the weak, broad feature at 10.7 \micron from PAH cations
\citep{Hony2001} are not clear in the original spectrum, but stand out
in the residual. In addition, we find that the profiles of the already
prominent PAH features become clearer when the \hoh emission is
removed. 

\begin{figure*}[t!]
\centering
\includegraphics[width=1\textwidth]{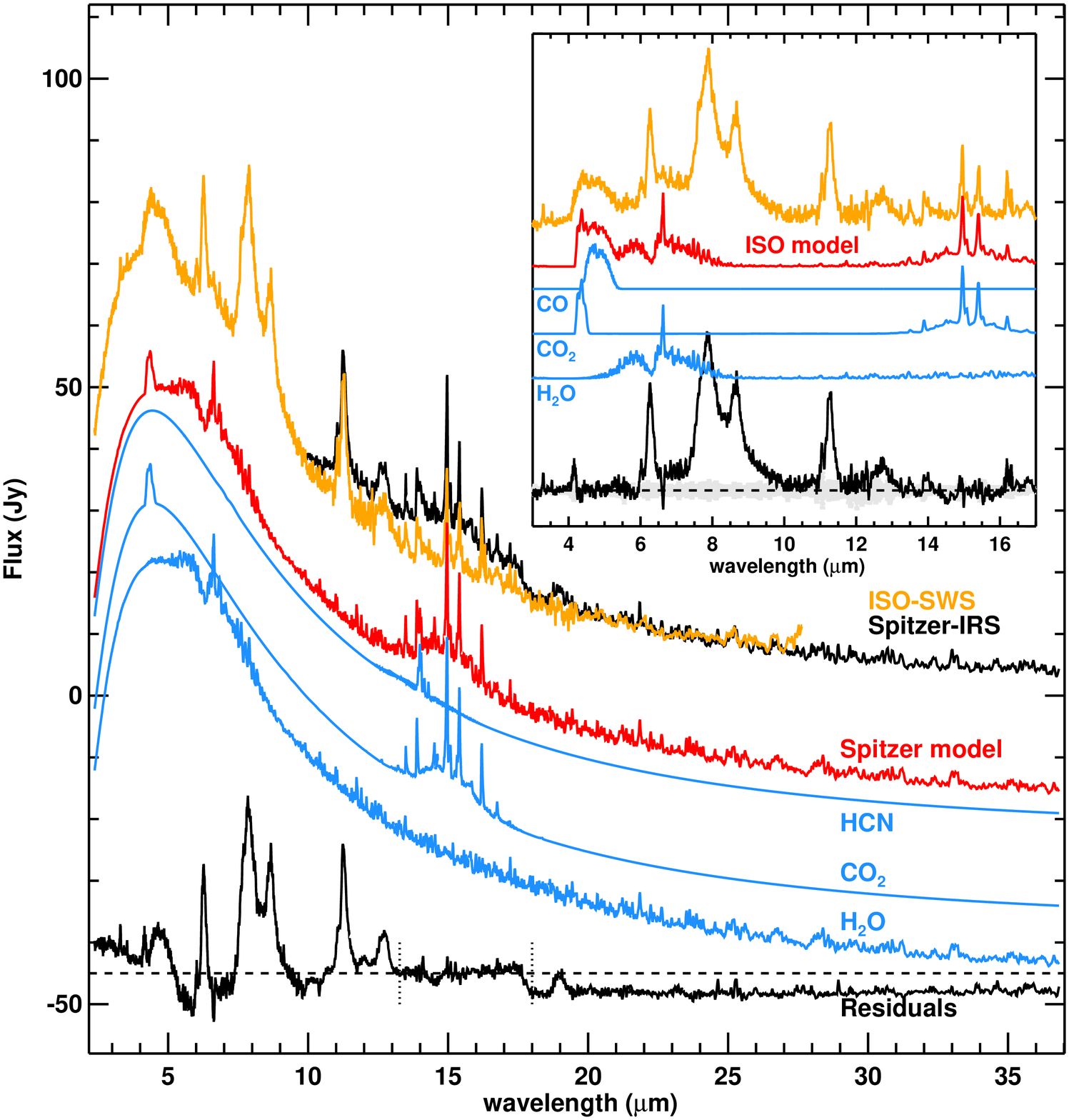}
\caption{Main plot: Spitzer-IRS spectrum (black) and ISO-SWS spectrum
  (light grey or orange in the online version) for HR 4049 with the
  best fit model (dark gray or red in the online version) to the
  Spitzer-IRS spectrum and models for the individual components
  for \coo, \hoh, and
  HCN offset below the spectrum (gray or blue in the online
  version) as well as the residuals (black).  The vertical dotted lines
  indicate the region of the spectrum used to fit the model to the
  Spitzer-IRS data. The model outside of these wavelength ranges
  (including the ISO-SWS spectrum) is the prediction based on these
  best fit parameters. Inset: Model fit to the ISO-SWS spectrum, with
  the best fit model (dark gray or red in the online version) and
  individual components for CO, \coo and \hoh offset below the
  spectrum (gray or blue in the online version). The residuals for
  this model are shown in black and the uncertainties in the data are
  shown in light gray.}
\label{fig:midir_fits}
\end{figure*}

\subsection{Residual Features}

There are a few interesting features in the residual spectrum. For
instance, we note a small absorption feature at 10.38~\micron. This
feature was previously observed in the carbon-rich pre-planetary nebula
\object{SMP LMC 11} by \citet{Malek2012}, who suggested that this
feature is molecular in nature, but they were unable to identify the
carrier. 

As described above, a few of the residual features in the LH spectrum
may be due to OH emission. However, we also observe an emission feature
at 25.99~\micron which could be due to  a fine-structure line from
[\ion{Fe}{2}] (see Fig.~\ref{fig:spitzeroh}). 
To confirm that this feature is real, we examined the background
observations and determined that this region of the spectrum was
featureless. Then we examined both nod positions in the
background-subtracted spectrum and noted that it appeared in both. 

This transition is from the J=7/2 to J=9/2 (ground) level in the
a$^6$D state. The next transition would appear at 35.35~\micron (J=5/2
to J=7/2), but we do not observe a clear feature from this line.
Instead, we observe a large and broad spike which appears to be from
noise (S/N decreases toward the end of the LH spectrum). If we
assume that the [\ion{Fe}{2}] is at the same temperature as the
molecular layer (500~K), we can estimate the
strength of the expected emission
at 35.35~\micron. Using the method described in
\citet{Justtanont1999}, we find that the 35.35~\micron line should be
80\% as strong as the 25.99~\micron line, which could be hidden by 
this spike in the noise. 

\paragraph{Plateau Emission}

In addition to the small emission features in our residuals, we note
that we are unable to properly reproduce the continuum beyond
$\sim$17.6~\micron. Indeed, examining Figs.~\ref{fig:sh_zoom} and
\ref{fig:midir_fits}, we see that we systematically overestimate the
continuum emission at LH wavelengths. As well, we note that one of the
major differences between the Spitzer-IRS and ISO-SWS spectra in
Fig.~\ref{fig:midobservations} is the presence of emission from a
continuum-like ``plateau" under the \coo emission in the Spitzer
spectrum. 

If we force the dust continuum to represent the LH continuum
more accurately, we find that our models cease to reproduce the narrow
emission bands from \coo. If we instead fit the narrow emission
features, we overestimate the dust continuum at LH wavelengths. We
have chosen to use a model which reproduces the narrow \coo features,
but we note that the dust continuum from our best fit model is
much higher than it should be as a result. 

Part of this plateau emission is likely due to the presence of
C$_{60}$. However, since the residual flux is much wider than the
17.4~\micron C$_{60}$ feature, this will not account for all of this
plateau. In addition, the carrier of this plateau emission appears
to have formed between the observations by ISO and Spitzer since this
is perhaps the most obvious difference between the two spectra. 
We therefore considered two possible sources for this plateau emission.

First, we considered the possibility that this could be due to broad PAH
emission from C-C-C bending modes, which has been observed to form a
continuum between 15 and 20 \micron \citep[e.g.~][]{vanKerckhoven2000,
Boersma2010}. These PAH plateaus also often show narrow features at
16.4, 17.4 and 18.9 \micron along with a weak 15.8~\micron feature
\citep{Tielens1999, Moutou2000}.  However, we do not see any obvious
 PAH features to the red of the 12.7~\micron feature. 
Therefore, we consider it unlikely that this plateau is due
to PAH emission.

Another possible candidate for this plateau emission is \coo. 
While our \coo models predict some continuum-like emission, it is not
enough to match the plateau we observe. We do find high optical depths
for \coo ($\tau_{\rm max} \approx 160$ for the main isotopologue using
a line width of 3 km s$^{-1}$, this would decrease to $\tau_{\rm max}
\approx 52$ for 10~km~s$^{-1}$) as well as some evidence for
temperature stratification in the residuals and possibly recent \coo
formation which could result in some non-LTE emission. We therefore
consider that the plateau emission could be due to \coo emission that
our models are unable to fit properly. We will explore this
possibility further in Section \ref{sec:discussion:time}.
We note that a similar plateau was also observed in
\object{IRAS 06338} \citep{Gielen2011} and was attributed to optically
thick \coo emission. 

\subsection{ISO Spectrum}
\label{sec:spitzerviso}

\coo, \hoh and HCN potentially all have features at the shorter
wavelengths covered by ISO-SWS in addition to the features we observe
in the Spitzer-IRS spectrum. We thus decided to compare the
predictions from our model at these wavelengths to the ISO-SWS
spectrum. We extended our model spectra to shorter wavelengths and
smoothed the spectrum covered only by the ISO-SWS spectrum to a
resolution of 300 to match these data. We present our prediction for
these bands in the main plot of Fig.~\ref{fig:midir_fits}.

These predictions do not appear to fit the ISO-SWS spectrum on first
sight. However, some of the features do appear to be reproduced. 
Examining the \coo feature at 4.2~\micron, we note that this band
is heavily blended with the much broader CO emission band at
4.6~\micron. It appears, however, that the \coo feature within this
broader band is reproduced reasonably well. 

If we examine our predictions for the \hoh spectrum, there is a broad
emission feature between 5 and 8~\micron. It is difficult to assess how
well this broad band reproduces the spectrum in these regions due to
the presence of the CO emission feature and the PAH features. However,
our model predicts a fairly strong \hoh feature at 6.62~\micron which
is much weaker in the ISO-SWS spectrum. 

Finally, although HCN also has an overtone mode at $\sim$7
\micron, our models do not predict it at these temperatures and column
densities. 

\medskip

Since the ISO-SWS spectrum has a much larger wavelength coverage, it
covers additional molecular bands (such as the 4.2~\micron \coo
feature) and species (CO at 4.6~\micron), so we decided to fit this
spectrum as well. We used the same techniques described for fitting
the Spitzer-IRS spectrum, but we included CO in our model
\citep[using line lists from ][ including the $^{13}$CO, C$^{17}$O and
C$^{18}$O isotopologues]{Goorvitch1994} and excluded HCN. We fit the
spectrum in several regions to include as many molecular features as
possible while excluding features we cannot fit (e.g.~the PAHs). Thus,
we used three fitting regions, first, between 2.4 and 5.8~\micron,
then from 9.5 to 10.5~\micron and finally from 14 to 22~\micron. 

The parameters for our best fit are presented in Table
\ref{table:mirfits} and this model is compared to
the 3 to 17 \micron region of the ISO-SWS spectrum in the inset of
Fig.~\ref{fig:midir_fits}. In our direct fit to the ISO-SWS spectrum,
we find that the fit in the 4~\micron region is greatly improved with
the addition of CO and a more appropriate scaling factor for the
continuum. However, we note that there are still issues in the \hoh
emission at 6.62~\micron where we fit a large emission spike which
does not appear in the spectrum. In addition, there are some residuals
at $\sim$16~\micron.
  
\paragraph{Time dependence}When we compare the ISO-SWS and Spitzer
observations, we find that there is a dramatic increase in emission
between 10 and 18~\micron. If we subtract a 1150~K black body
continuum from each spectrum and integrate the remaining flux between
13.4 and 16.8~$\mu$m, we find that the flux from molecular emission
has more than doubled in this region between the time of the ISO and
Spitzer observations (from $\sim2 \times 10^{-10}$~W~m$^2$ to $\sim5
\times 10^{-10}$~W~m$^2$). 

\section{Discussion}
\label{sec:discussion}

The better quality of the Spitzer-IRS data and current line lists
allow a much more in-depth analysis of the gas than was previously
possible. As we shall see, many of the results of this analysis have
important consequences for the properties of the circumbinary disk.

\subsection{Isotopic Ratios}

We remind the reader that in the optically thin limit,
\citet{Cami2001} determined that HR~4049 was extremely enriched
in \isotope[17]{O} and \isotope[18]{O} (\isotope[16]{O}/\isotope[17]{O}
= $8.3 \pm 2.3$; \isotope[16]{O}/\isotope[18]{O} = $6.9 \pm 0.9$).
In a subsequent study of the near-IR CO bands, \citet{Hinkle2007} 
found isotopic ratios consistent with solar values and they suggested
that the ratios determined by \citet{Cami2001} were incorrect because 
the \coo emission in HR~4049 is actually optically thick. Based on our
models, we are able to confirm the suggestion by \citet{Hinkle2007}
that the \coo is optically thick, but we  determine a ratio of
160$^{+90}_{-35}$ for \isotope[16]{O}/\isotope[17]{O} and
160$^{+40}_{-60}$ for \isotope[16]{O}/\isotope[18]{O}, indicating some 
enrichment in \isotope[17]{O} and \isotope[18]{O} relative to solar
values (\isotope[16]{O}/\isotope[17]{O} = 2700,
\isotope[16]{O}/\isotope[18]{O} = 479, \citet{Anders1989, Scott2006}).
In addition, these isotopes are enriched relative to tyical AGB stars
\citep[where both are on the order of a 10$^2$ to
10$^3$, see ][]{Harris1984,Harris1985,Harris1987,Harris1988,Smith1990}.
Indeed, the enrichment of \isotope[17]{O} and \isotope[18]{O} in
HR~4049 appears to be similar to that of the Ba star HD~101013
\citep[$^{16}$O/$^{17}$O = 100$^{+100}_{-50}$, $^{16}$O/$^{18}$O =
60$^{+100}_{-30}$, ][]{Harris1985a}. 

There are also recent detections of OC$^{18}$O in two other
binary post-AGB objects: \object{EP Lyr} and \object{HD 52961} which
both show enrichment in \isotope[18]{O}
\citep[$^{16}$O/$^{18}$O of 19 for EP Lyr and 100 for HD
52961,][]{Gielen2009}. As well as a study of \isotope[18]{O}
enrichment in R Coronae Borealis stars in which
\isotope[16]{O}/\isotope[18]{O} ratios less than one were found
\citep{Clayton2007}. 

We also note that the values we find here for the enrichment of
\isotope[17]{O} and \isotope[18]{O} are more suitable to the scenario
proposed by \citet{Lugaro2005} who suggested that the isotopic ratios
for oxygen in HR~4049 could be due to nova nucleosynthesis; though
HR~4049 lacks the UV flux for a typical white dwarf companion
\citep{Monier1999}. However, due to the high optical depths we find,
our values are poorly determined and our uncertainties on these values
are likely to be larger than we find from our models. We also suggest
that there may be similar issues with the
\isotope[16]{O}/\isotope[18]{O} values in the other post-AGB objects
for the same reason. 

\medskip

Our best fit models appear to indicate an enrichment in
\isotope[13]{C}, with \isotope[12]{C}/\isotope[13]{C} of
6$^{+2}_{-1}$. This relatively low value agrees with an
early termination of the AGB due to accelerated mass loss along the
orbital plane of the binary \citep{Iben1993}. This has been observed
in other binary post-AGB objects such as EP Lyr \citep{Gielen2008}.
However, optically thick \coo emission also makes this
ratio uncertain. 

\subsection{Gas Distribution and Disk Structure}

From our models, we find that not only is the \coo emission optically
thick, but the gas is optically thick across the entire spectrum
(e.g.~$\tau > 10$ for \hoh across the entire Spitzer-IRS spectrum).
This has some very important ramifications. For instance, when the gas
is optically thick, the flux will scale with the emitting area, thus
we are able to estimate the spatial extent of the gas using our
models. 

When we calculate our models, we obtain a scale factor ($f$) which
relates $I_\nu$ from our radiative transfer calculations to $F_\nu$.
We are able to relate this scale factor to the surface area ($A$) of
the emitting layer such that $f = A / 4 D^2$, where $D$ is the
distance to HR~4049.

From our model fit to the Spitzer-IRS data, we find a value of $1.60
\times 10^9$ for $f$. If we use a distance of 640~pc for HR 4049 \citep[as
described by ][]{Acke2013}, this scale factor corresponds to a
projected area of 1117~AU$^2$. If we perform the same analysis using
our best fit to the ISO-SWS spectrum (with $f = 1.57 \times 10^9$), we
find a projected area of 1097 AU$^2$, which agrees reasonably well
with the emitting region we estimate from our Spitzer-IRS data. Since
the scale factors we determine have not changed much between the ISO and
Spitzer observations despite the differences in column densities
between the two models, this also supports our claim of optically thick
gas. 

If we then assume an inclination angle of 60$^\circ$ (which is agreed
upon by both current models for the disk), we find an actual emitting
surface of 1290 AU$^2$ for our molecular emission. If we consider how
this surface area would fit into the current disk models, we note that
this would not fit on the inner rim of the wall model. 

This suggests that the gas belongs to a radially extended disk
instead. If we assume that the gas exists some distance from the
center of the binary system, we can estimate the maximum extent of the
disk. Were the gas to originate 10 AU from the binary \citep[as the
dust in the wall model,][]{Dominik2003}, it would extend to a distance
of 23 AU. If the gas begins at 15 AU from the binary \citep[based on
the interferometric observations by][]{Acke2013}, it would extend to
25~AU, a distance which agrees well with the maximum radial extent of the
disk determined from the interferometric observations by
\citet{Acke2013}. 

\medskip

It would be reasonable to suppose that the gas is mixed in with any
dust in the circumbinary disk of HR~4049. As described by
\citet{Dominik2003}, dust grains in a gas-rich disk tend to
settle toward the midplane of the disk, so the gas we observe could form
a sort of atmosphere on the outside of the disk. There could also be
more gas inside the disk which we are unable to observe since this
atmosphere is optically thick. However, the gas on the inside of the
disk could contribute to the opacity of the disk by providing some
continuum-like emission similar to that we observe in our models. 

As the LTE models reproduce the emission spectrum
well and the gas appears to be reasonably warm (500~K), we find
that we cannot reconcile our observations to a wall-type model for the
dust. The wall model contains cold dust beyond the inner rim of
the disk \citep{Dominik2003}, which would result in cold gas in this
region. Instead, we observe a large region of warm
gas so we find this sort of model improbable.  

It would also be difficult to reconcile our observations to the
disk model presented by \citet{Acke2013} in which the dust is
optically thin. \citet{Acke2013} include gas in their disk to determine
the scale height, however \coo and \hoh are both excellent at trapping
infrared radiation, especially at high column densities like those we
observe in this disk and these effects are not included in their model. 

Since these molecules are largely transparent at optical wavelengths,
the stellar radiation will warm the dust grains. These grains will
then re-emit this radiation in the infrared which will be absorbed and
re-emitted by the gas in the disk, effectively trapping the radiation
inside the disk. This will not only have the effect of warming the
disk overall, but it will also keep the temperature relatively
homogenous throughout the disk. Thus the disk could have the sort of
narrow temperature range suggested by \citet{Dominik2003}, who
determined that the SED of HR~4049 can be fit either by a 1150~K black
body or by the sum of several equally weighted black bodies within a
range of 880 K $\le T \le$ 1325 K. 

However, in LTE, a gas cannot appear in emission in front of a hotter
background source. Thus, we will explore alternate excitation
mechanisms which could result in bands which emit in a way which
appears similar to LTE in Section \ref{sec:discussion:time}. 

\subsection{Total Gas Mass}

Using the projected size for the emitting region and our column
densities, we can estimate lower limits for the masses of the
molecular species we 
observe in this system. Then, combining these with photospheric
abundances, we can estimate a lower limit for the total mass of the gas in
the circumbinary environment of HR~4049.

From our model fit to the Spitzer-IRS data, we calculate a mass of 
$9.19 \times 10^{-8}$ M$_\sun$ for \coo, $1.50 \times 10^{-5}$ M$_\sun$
for \hoh and $3.56 \times 10^{-9}$ M$_\sun$ for HCN. Similarly, for
the ISO-SWS model, we find masses of $5.69 \times 10^{-9}$ M$_\sun$
for \coo, $9.28 \times 10^{-8}$ M$_\sun$ for \hoh and $5.74 \times
10^{-5}$ M$_\sun$ of CO. 

Since the gas is oxygen-rich, we use the carbon abundance
determined by \citet[log $N_C/N_H = 8.41 - 12$]{Waelkens1991} along
with the number of carbon-containing molecules and determine a lower
limit for the total gas mass of $7.98 \times 10^{-3}$ M$_{\sun}$ in
the disk. This estimate is higher than the mass estimated by
\citet{Dominik2003}, who estimated $2.85 \times 10^{-4}$ M$_\sun$ for
the total mass of the disk. This could indicate a higher gas to dust
ratio in the disk (they use a value of 100) or the presence of even
more dust in this system than predicted by the wall model. If we
compare our estimate for the gas mass to the dust mass from \citet[in
which $M_{\rm dust} = (1.0 \pm 0.4) \times 10^{-8}$
M$_\sun$]{Acke2013}, we find a gas-to-dust ratio of approximately
10$^6$. This appears unreasonably high, however this is the
estimated mass for only the small grains in the disk, which are
responsible for the near- and mid-IR SED as well as the optical and UV
extinction. Their model also includes a cold dust component to
reproduce the flux at the far-IR and submm wavelengths, which would contain
more mass.

Since the gas is optically thick, this estimate will not include
all the gas in the system (e.g.~any gas beyond the optically thick
layer or on the side of the disk inclined away from us is not
included). In addition, since the majority of the gas included in our
estimate is CO which only appears in the ISO-SWS spectrum, this may
not describe the current gas mass since, as we will describe
presently, it appears there is significant ongoing gas formation in HR
4049.

\subsection{Time Evolution}
\label{sec:discussion:time}

It is somewhat surprising to see that the \coo flux has increased by a
factor of 2.5 between the ISO-SWS and Spitzer-IRS observations (see
Section \ref{sec:spitzerviso}).
Since we scaled the Spitzer-IRS observations to match the IRAS flux
point at 25~\micron, we considered that this could be an issue in our
comparison of these observations. However, the ISO-SWS spectrum was
also scaled to the same point so this is unlikely to change our
observation. Furthermore, it is not just the flux that has
changed under the \coo emission features, but also the shape of the
``continuum" and features in this region. Thus, it appears that the
change in the emission features is real.  

We compared the phase between the two observations using the
phase information from \citet{Bakker1998}. The SH observations were
taken at a phase ($\phi$) of 0.844, while the ISO-SWS speed 2
observations were taken at $\phi = 0.041$, near the photometric
minumum. There are also the ISO-SWS speed 1 observations available,
which were taken at a similar phase as the Spitzer-IRS
observations ($\phi = 0.738$).

We thus compared the two ISO-SWS observations to see if there
was a change in this region of the spectrum which could be attributed
to phase. When we did so, we found
that the spectra of the speed 1 and speed 2 ISO-SWS observations were
roughly the same within the uncertainty mesurements on the fluxes. 

Thus, we conclude that the amount of emitting \coo gas we can observe has
increased between the observations by
ISO-SWS and those by Spitzer-IRS observations. If the gas were
optically thin, this would imply that CO$_2$ has been forming at a
rate of $3.68 \times 10^{-9}$ M$_\sun$ yr$^{-1}$ assuming a constant
formation rate. We note that this represents a considerable and
rapid increase in \coo in the system, however, as we
discovered, the \coo emission in the
mid-IR spectrum is not optically thin so this represents a lower limit
to the increase in \coo. In addition, we cannot know whether the
formation has been continuous between the observations so this is a
crude estimate for the formation rate. 

%\medskip

While \citet{Bakker1996} reported a mass-loss rate from HR~4049 of $(6
\pm 4) \times 10^{-7}$ M$_\sun$ yr$^{-1}$, the amount of carbon and
oxygen in the stellar winds is insufficient to permit the formation of
so much new material. Therefore, we consider that the \coo we observe
is forming from the interaction between the stellar wind and the dust
disk. 

The ongoing formation of oxygen-rich gas thus suggests that the dust 
contains an oxygen-rich component (e.g.~silicates) and the absence of
features from these dust species in the spectrum would therefore be
due to obscuration of these features by optically thick gas; or by the
dust being optically thick or perhaps　being composed of large dust
grains which have a smooth opacity. If the dust is primarily
oxygen-rich, then HR~4049 would also be consistent with other post-AGB
binaries in this regard　\citep[e.g.~][]{Gielen2011a}. 

While this may appear to be inconsistent with the fact that small
dust grains are required to explain the optical and UV extinction, it
was noted by \citet{Dominik2003} that the extinction at short
wavelengths and the IR excess do not need to be caused by the same
population of dust grains. Thus, although the optical and UV
extinction is described well by a population of small grains of
amorphous carbon or metallic iron as described by \citet{Acke2013},
these grains cannot be the only dust component in the disk since they
cannot allow the formation of the oxygen-rich gas we observe here. 

The presence of small carbonaceous grains in the upper region of
the disk could thus contribute to the extinction at short wavelengths
while a population of oxygen-rich grains could contribute to part of
the IR emission. 

\medskip

The evidence for ongoing formation of gas in the circumbinary disk
of HR 4049 also suggests that the molecules are not actually in
LTE, and the emission we observe may be due to pumping---either
{\em radiative} pumping or {\em formation} pumping. In both cases,
pumping results in a large fraction of molecules in highly excited
vibrational states, either through absorption of higher energy
photons (near-IR or even UV) or alternatively as a direct result
of the formation of the molecules which leaves them in the excited
states. After being pumped, they cascade down by 
emission of the IR photons we observe. Pumping would thus always
produce emission at mid-IR wavelengths even when in front of hot
dust. Superficially, such emission spectra could resemble LTE
models; their main difference would be the presence of many bands
from higher vibrational levels. This could help explain the
plateau emission as a forest of \coo emission lines from higher
energies.

It is thus possible that the \coo and \hoh are being
radiatively pumped the hot dust, but this has not been
directly observed. Note that all these species have
strong electronic absorption bands at UV wavelengths which could
be the source of the pumping through absorption of stellar
radiation as well. Given the high column densities of these species, it
would be interesting to investigate whether this could be a
contributing source of the observed UV deficit in HR 4049
\citep{Lamers1986}. Formation pumping is certainly an appealing
alternative, but is not well studied. While models exist for the
formation pumping of H$_2$ after formation on dust grain surfaces
\citep[e.g.~][]{Gough1996, Takahashi2001}, but to our
knowledge no such models exist for \coo or \hoh.

\medskip

Finally, we note that while it appears that gas has been forming in
the disk of HR~4049, the overall emitting surface does not seem to
have changed significantly (remaining at $\sim$1300 AU$^2$). This
agrees very well with the idea that the disk is relatively stable and
long-lived (also supported by the unchanging CO overtone absorption
observed between the observations of \citet{Lambert1988} and those by
\citet{Hinkle2007}). 

\citet{Dominik2003} described how a gas-rich disk (such as the one we
describe) would tend to settle toward the midplane and expand outward.
Using a gas-to-dust ratio of 100 and an initial scale height of 4 AU,
they estimated that the scale height of the disk would decrease by
half in 150 years. This effect could be mitigated by a higher
gas-to-dust ratio and we suggest that the gas-to-dust ratio is likely
to be greater than 100 in this environment. Indeed, if the dust is
being slowly destroyed by stellar wind and gas is being formed, this
ratio is likely to be increasing.

\section{Conclusion}
\label{sec:conclusion}

The Spitzer-IRS observations clearly reveal that the molecular gas
in the circumbinary disk of HR 4049 is optically thick at infrared
wavelengths and that the emission originates from a radially
extended disk. The gas causes a strong greenhouse effect that plays
a significant role in determining the thermal structure in the
disk. Including the effect of optical depth, we determine that there
is less of an enrichment in $^{17}$O and $^{18}$O than previously
reported. Additionally, changes in the observed flux between ISO and
Spitzer observations suggest ongoing chemical processing of
oxygen-rich dust.

\acknowledgements
We acknowledge the support from the Natural Sciences and Engineering
Research Council of Canada (NSERC).
This work is based (in part) on observations made with the Spitzer
Space Telescope, which is operated by the Jet Propulsion Laboratory,
California Institute of Technology under a contract with NASA.
This research ha salso made use of NASA's Astrophysics Data System
Bibliographic Services and the SIMBAD database, operated at CDS,
Strasbourg, France.

%\bibliographystyle{NBaa}
%\bibliography{library}
%\bibliography{../../library}{}

%\end{document}

\end{document}